\documentclass[prl,aps,12pt
]{revtex4}
\usepackage{amsmath,amssymb,graphicx,mathrsfs,hyperref,slashed}
\newcommand{\be}{\begin{equation}}
\newcommand{\ee}{\end{equation}}
\newcommand{\bea}{\begin{eqnarray}}
\newcommand{\eea}{\end{eqnarray}}

\def\({\left(} \def\){\right)}

\begin{document}
\title{
\vspace{3mm}
\vspace{0.3cm}
{Resisting collapse: How matter inside a black hole can  withstand gravity}}
\author{ Ram Brustein${}^{(1)}$,  A.J.M. Medved${}^{(2,3)}$
\\
(1)\ Department of Physics, Ben-Gurion University,
    Beer-Sheva 84105, Israel
(2)\ Department of Physics \& Electronics, Rhodes University,
  Grahamstown 6140, South Africa \\
(3)\ National Institute for Theoretical Physics (NITheP), Western Cape 7602,
South Africa \\
    ramyb@bgu.ac.il,\  j.medved@ru.ac.za }
\begin{abstract}
How can a Schwarzschild-sized matter system avoid a fate of gravitational collapse? To address this question, we critically reexamine the arguments that led to the ``Buchdahl bound'', which implies that  the minimal size of a stable, compact object must be larger than nine eighths of its own Schwarzschild radius. Following Mazur and Mottola, and in line with other counterexamples to  the singularity theorems, we identify large negative radial pressure extending to the gravitational radius as the essential ingredient  for evading the Buchdahl bound.  Our results are novel  although  consistent with many other investigations of models of non-singular black holes. It is shown in particular  that a large negative pressure in the framework of classical GR translates into a large positive pressure once quantum physics is incorporated. In this way, a Schwarzschild-sized bound state of closed,  interacting fundamental strings in its high-temperature Hagedorn phase can appear to have negative pressure  and thus the ability to resist gravitational collapse.

\end{abstract}
\maketitle

\section{Introduction}

The tension between black hole (BH) evaporation  and the rules of quantum mechanics \cite{info,Sunny,Mathur1,Braun,AMPS,MP,Mathur4}, as well as the recent discovery of gravitational waves being emitted from colliding BHs and neutron stars, has reignited  interest in the following question: What is the final state of matter after it  collapses to form  a BH?

The singularity theorems of Hawking and Penrose \cite{PenHawk1,PenHawk2} decree that the final state of collapsing matter, when considered within the purview of classical general relativity (GR), must be singular.  An elegant discussion which preceded these theorems  was provided in a simplified context by Buchdahl \cite{Buchdahl} (and later  by Chandrasekhar \cite{chand1,chand2} and  Bondi \cite{bondi}).  Buchdahl was able to show that a ``conventional" matter system cannot be stable within its own Schwarzschild radius. In fact, the ``Buchdahl bound'' on the outermost radius of a stable fluid sphere $R$ is somewhat larger than the Schwarzschild limit,
$\;R\geq \frac{9}{8} R_S\;$ \cite{Buchdahl}.
This result puts a damper on the idea that an ultra-compact object could play the role of a BH while being fundamentally different from the BHs of GR.

Buchdahl  invoked the usual assumptions that lead to a Schwarzschild geometry. He further assumed both causality and a classical energy condition (the strong energy condition), as similarly required by the singularity theorems. These assumption were implemented by requiring that both the energy density and the pressure are positive. Regarding the matter distribution, Buchdahl made three  additional assumptions. As more recently discussed in \cite{MM}, these are  (I) the isotropy and positivity of the pressure,   (II) the monotonic dilution of the matter distribution  when  moving  outward from the  center and   (III) the continuity of the time--time component of the Schwarzschild metric and its first derivative  across the boundary of the matter sphere. The rest of the proof relies only upon the Einstein field equations of GR.

The conclusion is that any reasonable BH substitute --- meaning  an
ultra-compact object which  resembles a BH --- has to evade the Buchdahl bound by invalidating  at least one of its assumptions.

Following to some extent \cite{MM}, our description of matter that can avoid gravitational collapse begins with a discussion about its  pressure. We will, in particular, argue that (at least) the radial component of the pressure  needs to be negative and large in magnitude throughout the entire interior.  Such matter has to be perceived to  have a solid  composition  and  be  under severe tension, as noticed long ago by Bondi \cite{bondi}.  Matter with negative pressure  is important on three fronts:
(a) It  can explain why a system under collapse does not have to  comply with the singularity theorems. This follows in  analogy to the  negative pressure of the cosmological dark energy.
(b) Any  negative component of pressure would lower an object's gravitational or Komar mass density \cite{Komar}  and thus can  impede  collapse by weakening the gravitational field.
(c) A  negative radial pressure invalidates not only assumption~(I) from  the prior list  but also Buchdahl's assumption  in (III) regarding the continuity
of the first derivative of the metric.

We will be discussing a new model for the BH interior but eventually reveal its ``true'' identity; namely, our recently proposed
``collapsed polymer'' model \cite{strungout} as viewed from a classical perspective.  But some other models which similarly utilize a large negative pressure as means for inhibiting collapse were proposed. The typical phenomenological description of a non-singular BH  uses a classical metric  and introduces a new length-scale parameter, say $L$. It is  assumed that the metric deviates from that of classical GR when the curvatures becomes of order $1/L^2$ in the core of the regular BH.   The deviations are supposed to imitate some quantum-gravity corrections to classical GR and should result in a singularity-free solution. The first model of a non-singular BH was proposed by Bardeen \cite{bardeen}, who considered a charged matter core inside the BH instead of a singularity. A model of a non-singular evaporating BH  was later discussed in \cite{FrVi,hayward}. Many more references and detailed reviews of additional models of regular BHs can be found in \cite{splucc,frolovrev}. The ``gravastar'' model \cite{MMfirst}  introduces negative pressure in terms of  a de Sitter interior, with a suitable outer layer of matter to ensure stability. The ``black star'' model \cite{barcelo} is based on the  notion that  high-energy physics could provide a mechanism for delaying gravitational collapse until the
  back-reaction effects of the matter fields  have had sufficient time to modify the semiclassical picture.  Recently, Carballo-Rubio implemented the effects of the back-reaction and obtained a non-perturbative (but still semiclassical) solution  that shares some features in common with both the gravastar and black star \cite{CR}. In \cite{frolov,visser},  it was shown, generically, that regular BH solutions are not self-consistent, as  the energy emitted by them during  evaporation can be much larger than their original mass.  Our current model of interest (and therefore the polymer BH) can be viewed as a limiting case of the black star model.
 The essential difference is that, in our case, the outermost surface of the collapsing matter reaches the Schwarzschild radius before stabilizing. Our model is also consistent with the recent discussion in \cite{chingaling}.

Let us now call upon an analysis  in  \cite{MVM}. There it was established that  the radial pressure $p_r$ of any static matter just outside of a Schwarzschild  horizon has to satisfy the condition
$\;p_r=-\rho\;$, with $\rho$ being the energy density. This condition is necessary  to ensure the near-horizon regularity of  curvature  invariants like  the  Kretschmann scalar and,  by continuity,  has to hold  all the way up to and including the horizon.  The transverse components of pressure $p_{\perp}$,  on the other hand, cannot  be similarly constrained by appealing to regularity, as
was  also clarified  in  \cite{MVM}. One way to enforce the condition $\;p_r=-\rho\;$ at the horizon is to insist  that it be true throughout the interior; for instance, by replacing the GR description of the BH interior with a ball composed of some matter under tension.   But what type of matter
can satisfy   $\;p_r=-\rho\;$ and, moreover, find its way  into
the interiors of BH-like, ultra-compact objects? 
The answers to these puzzles will be presented later on in the paper when we consider a different perspective of the very same model.

\section{Evading the Buchdahl bound with negative pressure }

Let us start by specifying  the  geometry and matter for the inside of a spherical, ultra-compact  object of radius $R$.  We assume  a ``Schwarzschild-like'' geometry for the interior (whereas that of the exterior is   precisely  Schwarzschild),~\footnote{Throughout the paper, we assume a Schwarzschild geometry in the exterior of the ultra-compact object and focus primarily on the case of $\;D=3+1\;$ spacetime dimensions. This choice of $D$ is made for the sake of clarity  as none of our conclusions change for  $\;D>4\;$. Fundamental constants
and numerical factors are often neglected unless needed to establish an exact equality and not just a scaling relation. A prime on a function denotes a radial
derivative.}

\be
ds^2\; =\; -f(r) dt^2 + \frac{1}{{\widetilde f}(r)} dr^2 + r^2 (d\theta^2+ sin^2 \theta d\phi^2)\;.
\ee
 It is further assumed that $\;p_r=-\rho\;$,  as motivated  in the previous section,  which translates into  an  energy--momentum--stress (EMS) tensor of the form
\be
\label{polytmunu}
T^\mu_{~\nu}\;=\; \left(
             \begin{array}{cccc}
               -\rho & 0 & 0 & 0 \\
               0 & -\rho & 0 & 0 \\
               0 & 0 & p_{\perp} & 0 \\
               0 & 0 & 0 & p_{\perp} \\
             \end{array}
           \right)\;.
\ee

For this setup, the Einstein equations
\begin{align}
\label{E1}
\left(r{\widetilde f}\right)'  & \;=\; 1-8\pi G r^2 \rho \;, \\
\label{E2}
\left(rf\right)''\; & \;=\; 16\pi G r p_{\perp}\;,
\end{align}
can be combined into
\be
\label{E3}
\left(\rho r^2\right)'  \;=\; - 2 r p_{\perp}\;,
\ee
which is guaranteed by the covariant conservation of the EMS tensor.

Defining
\be
\label{mofr}
m(r)\;=\;4\pi \int\limits_0^r dx\, x^2 \rho(x)\;
\ee
and taking into account that $\;m(R)=M\;$ is the total mass of the ultra-compact object, one finds that solving
 Eqs.~(\ref{E1}) and~(\ref{E2}) leads to
\be
f(r)\;=\; {\widetilde f}(r) \;=\; 1- \frac{2 G m(r)}{r}\;.
\ee
The value of $p_{\perp}$ is  determined by the function $m(r)$,  which is  determined in turn  by $\rho(r)$. Additionally, $\rho$ (and  likewise for  $p_r$ and $p_\perp$) has to vanish for $\;r\geq R\;$. According to Buchdahl, the function $f(r)$ and its first derivative are supposed to be  continuous across $\;r=R\;$.

It is instructive to consider some explicit examples that satisfy $\;p_r=-\rho\;$.  To this end, let us then parametrize $\;\rho=B r^\alpha\;$, where $B$ is a dimensional constant. A well-known example is the gravastar solution \cite{MMfirst}, for which $\;\alpha=0\;$ or $
\;\rho=const.\;$  In this case,  $\;p_{\perp}=-\rho\;$, so that the
interior of the ultra-compact  object,  $\;r\leq R\;$, is part of a de Sitter spacetime.  It follows that  $\;f(r)= 1- r^2/R^2\;$  on the inside, whereas
$\;f(r)= 1-2GM/r\;$ with $\;M=\frac{4}{3}\pi B R^3\;$ is the outer solution. It is clear that the function $f(r)$ is continuous across the outermost surface but its derivative is not. In \cite{MM},
this issue was resolved by adding a surface layer of matter with a surface tension that is determined by the continuity condition on $f'(r)$.

Another solution of this sort is found by setting $\;\alpha=-2\;$.  Although  a novel choice, it will be shown
later that this is really an alternative description of our ``collapsed polymer'' model \cite{inny,strungout,emerge}. In this case and for  $\;r\leq R\;$, then $\;f=0\;$ and
\bea
\label{polyrho}
r^2 \rho &=& \frac{1}{8 \pi G}\;, \\
\label{polypr}
r^2 p_r &=& - \frac{1}{8\pi G}\;, \\
\label{polypt}
p_\perp &=& 0\;,
\eea
with~\footnote{In spite of appearances, this is a non-singular matter distribution as the physically relevant quantity, $dr\;4\pi r^2 \rho$, is constant, finite and regular throughout. Put differently, the metric is regular throughout and the EMS tensor can be viewed as a consequence rather than a source of this  geometry.} the units chosen to yield   $\;m(r)=\frac{1}{2G} r\;$.
This solution  differs qualitatively from all other regular BH solutions by being null throughout the interior and  was obtained previously but in an
entirely  different context \cite{guendel1,guendel2}.  
Continuity across the outer surface  then requires  $\rho$ and $p_r$  to vanish  as  $\;r\to R\;$. This ensures that the function $m(r)$ approaches a constant
at the surface, which then  allows $\;f'(r)|_{r\to R}\sim 1/R^2\;$ on the inside to be   matched to its outer Schwarzschild value of $2MG/R^2$. Consistency further requires  the energy density of the interior matter to transition from a power law to zero in a smooth way as $\;r\to\;R$. This is accompanied by some positive transverse pressure which also vanishes smoothly in the same limit ({\em cf}, Eq.~(\ref{E3})) but, unlike the gravastar, no additional layer of matter is needed.

The vanishing of the metric function $f$ everywhere inside  is not a problem
because the Einstein equations, curvature tensors, curvature invariants and metric determinant are all as regular as they would be  at the horizon of a conventional Schwarzschild BH. However, as far as  solutions of GR go, this one does seem  rather peculiar. A heuristic way of understanding this solution goes as  follows: Let us recall that  the radial pressure of any form of matter just outside a Schwarzschild  horizon has to satisfy the condition $\;p_r=-\rho\;$ \cite{MVM}.
A  subtle consequence  of this constraint  is  that each  spherical slice of the solution has to behave  just like a horizon if its outer ``skin'' is peeled away. And yet, by Gauss' Law  along with  spherical symmetry, the  presence  of the outer skin is irrelevant to the inside. It  can then be concluded that $\;p_r=-\rho\;$ has to be  the {\em physical} radial pressure throughout the interior matter. The vanishing of the transverse pressure and the metric function then follow from the Einstein equations.

\section{Hydrodynamic equilibrium and stability}

In the previous section, it was shown that the solution in Eqs.~(\ref{polyrho})-(\ref{polypt}) can describe a Schwarzschild-sized, spherical mass distribution. We will now show, following Chandrasekhar \cite{chand1,chand2}, that such a distribution can be both in  hydrodynamic equilibrium  and   stable
against radial oscillations. In fact, it will be further shown that these oscillations are completely absent in the equilibrium state.

The relativistic hydrodynamic equations can be cast in the form of conservation equations,
\bea
\label{hydro1}
\partial_\mu\left( \sqrt{-g}\,\rho\,u^\mu \right) &=& 0\;, \\
\label{hydro2}
\partial_\mu\left( \sqrt{-g}~T^{\mu}_{~\nu} \right) &=& 0\;.
\eea
where  $\;u^\mu= (\gamma,\gamma\vec{\text v})$ is the 4-velocity of the fluid, $\vec{\text v}$ is the 3-velocity and  $\;\gamma=1/\sqrt{1-\vec{\text v}^2}\;$.

In our case, Eqs.~(\ref{hydro1}), (\ref{hydro2}) lead to three more equations,
\bea
\label{hydro3}
\partial_t(r^2~\rho~\gamma) + \partial_r(r^2~\rho~\gamma {\text v}_r) &=& 0\;, \\
\label{hydro4}
\partial_t\left( r^2 ~\rho\right)  &=& 0\;, \\
\label{hydro5}
\partial_r\left( r^2 ~\rho\right)  &=& 0\;.
\eea

Hydrodynamic equilibrium requires that the 4-velocity remains constant, from  which  $\;\gamma,\;{\text v}_r=const.\;$  follows as well.
Amazingly, each of the terms in Eqs.~(\ref{hydro3})-(\ref{hydro5}) then vanishes for hydrodynamic equilibrium  simply because $r^2\,\rho$ in Eq.~(\ref{polyrho}) is a temporal  and radial constant.

The stability analysis proceeds by perturbing the metric, the velocity and the
EMS tensor by small perturbations that depend on $r$ and $t$. The expansion parameter can be regarded as 
$\frac{\delta\rho(r,t)}{\rho_0}$. The full process is  described explicitly by Chandrasekhar in Sect.~{IV} of \cite{chand2}. This analysis is straightforward but quite long and technical and will  not  be
repeated here. Rather, we will briefly explain our results
while  referring to some expressions in
\cite{chand1}, which is a shorter companion paper to \cite{chand2}.

For our case with  $\;p_r+\rho=0\;$ and $\;g_{tt}=g^{rr}=0\;$ for the background solution, the perturbations to the off-diagonal elements of the EMS tensor are trivially vanishing at linear order, whereas the linear perturbations  $\delta \rho$,
$\delta g_{rr}$ and $\delta p_r$ vanish according to Eqs.~(10),~(6) and ~(7) in \cite{chand1}, respectively.
Let us recall these  expressions,
\bea
\delta \rho&=&-\left[r^2\left({p_r}_{~\!\!0}+\rho_0\right)\xi\right]'/r^2\;,  \\
\left(r g_0^{rr} \delta \lambda\right)' &=& -r^2 \delta \rho\;, \\
r^2 \delta p_r &=& g_0^{rr}\left(1+r\nu'_0\right)\delta\lambda
-rg^{rr}_0\delta \nu'\;,
\eea
 where $\;g_{rr}= e^{\lambda}\;$,  $\;g_{tt}= -e^{\nu}\;$,
a subscript $0$ indicates a background quantity and  $\xi$ is
defined implicitly through $\;{\text v}_r=\frac{\partial\xi}{\partial t}\;$.

The vanishing of $\delta g_{tt}$ is more subtle as $\delta \nu$  is always accompanied by a  factor of either $\; {p_r}_{~\!\!0}+\rho_0\;$ or $g_0^{rr}$. The easiest way to see that $\delta g_{tt}$  similarly vanishes is to impose that the determinant of the perturbed metric is regular, which necessitates that $\;g_{tt}=-g^{rr}\;$ to all orders. Alternatively, one could regularize the leading-order components by regarding $g_0^{rr}$ and ${g_{tt}}_{~\!\!0}$ as  small constants and take them to zero at the  end. One then finds that $\delta g_{tt}$ vanishes given that all the other perturbations do.

Now, since the perturbations are all vanishing and the velocity must be constant, there is no opportunity for radial oscillations to occur. One can
further verify this result by inspecting some additional equations in
\cite{chand1}. We conclude that the equilibrium configuration
of a spherical matter distribution  with $\;p_r=-\rho\;$ is indeed stable, at least to linear order. A second-order perturbative  analysis could, in principle, go either way.
However, as will be made clear in the section to follow, we have reason to believe that this stability would persist even at higher orders.

\section{Matter with negative pressure: the inside story}

We have presented arguments supporting the idea that matter with large negative radial pressure can resist gravitational collapse, even when it is confined to a Schwarzschild-sized region. In light of these arguments, it is worth  recalling the earlier-posed questions: What type of matter can  satisfy $\;p_r=-\rho\;$ and how can such  matter find its way inside of  BH-like, ultra-compact objects?

Our proposed answer  to either query is to suggest that such matter is something of an illusion and to rather  consider a different form of matter: A  bound state of fundamental closed, interacting strings, excited to temperatures just above the Hagedorn temperature; what we have been calling the collapsed polymer model \cite{strungout}. The model is described in detail in \cite{emerge} and we will review its main ingredients here for completeness. This proposal was inspired in part by \cite{SS,LT} and  motivated by the observation that the BH interior has to be in a strongly non-classical state \cite{density,inny,noclass}, even at times well before the Page time \cite{page}. Fundamental strings could be produced  out of  whatever form of  matter that collapses into a BH-like object, as long as the resulting  string state is entropically favored. This answers  the latter of the above questions, but addressing the former ({\em i.e.}, explaining how this form of matter connects back to the negative-pressure solution) will require some additional work.

The high-temperature phase of fundamental strings is called the Hagedorn phase \cite{FV,MT,AW,Deo,BV,HP}. In this phase, the spectrum consists of an exponentially large number of closely packed states and its canonical ensemble is subject to large fluctuations that diverge at the Hagedorn temperature. For such a state, the entropy $S$ is equal to the total length of the strings $L$  in units of the  string scale $l_s$,  $\;S=L/l_s\;$, and  the spatial configuration of the strings can be viewed as an $N$-step random walk with $\;N\sim S\;$. Free strings occupy a region in space whose linear size $R$  is the random-walk scale, $\;R\sim \sqrt{N}\;$, but  attractive interactions result in a smaller value of $R$  \cite{HP,DV}.
And so, for (attractively) interacting strings, one can expect that $\;R\sim N^\nu\;$ with $\;\nu\leq 1/2\;$ and, for  an area law  (as in the case of BHs), the condition becomes $\;\nu=1/(d-1)\;$ in
$\;D=d+1\;$ dimensions. This may be $\;\nu=1/2$\; when $\;D=4\;$, but the size of the random walk is still parametrically smaller than that of the free-string case, as now one finds  $\;R\sim g\sqrt{N}\;$
with $\;g<1\;$ being the strength of the string coupling.

The equation of state for this high-temperature string phase is famously $\;p=\rho\;$. In such a phase, in string units, $\;s=\sqrt{\rho}\;$ and  $\;1/T=ds/d\rho=s/2\rho\;$ which are, of course, consistent with the thermodynamic relation
$\;sT=p+\rho\;$. Here,  $s$ is the entropy density. Thus,  $s$ is as large as it can be in comparison  to $\rho$, implying  entropic dominance \cite{Brustein:1999md}.

How can the {\em positive} pressure $\;p_r=+\rho\;$ of stringy matter  be reconciled with the requirement that $\;p_r=-\rho\;$? Negative normal pressure, or tension, is a common phenomenon in solid matter and especially in polymers. It is then a natural tendency to expect the collapse-resistant matter to be composed of tensile material.  The purpose of this section is to explain that this tendency is unwarranted. If one ignores the entropy of matter, the relation $\;0=sT=p+\rho\;$ leads to $\;p=-\rho\;$.  Whereas, if the entropy is maximal and  $\;sT=2\rho\;$ is taken into account, then  $\;2\rho=sT=p+\rho\;$ implying $\;p=+\rho\;$. In short, what one perceives  as being  matter under tension turns out to be a completely different form of matter.

In classical GR, for $\;p_r=-\rho\;$, the collapse is avoided mechanically, while if $p_r=+\rho$ with maximal entropy, the collapse is avoided due to the following entropic consideration: The only way that the strings can collapse further and occupy a smaller region in space  is by  splitting up  into a collection of parametrically smaller strings. However, a configuration of many smaller strings is strongly disfavored in comparison to one long string because the entropy of latter is substantially larger.

The stringy perspective also provides us with an understanding of why the 
negative-pressure solution can be expected to be stable. The polymer model  is completely
stable in the absence of string interactions, which is itself an $\hbar$ effect.
Hence, the negative-pressure model, which is strictly classical by design,
should be similarly stable.

Let us briefly summarize the calculation of the radial pressure $p_r$ from the polymer model perspective. (We assume $\;p_{\perp}=0\;$ on the basis of spherical symmetry.) At equilibrium, the  energy density $\rho$ and entropy density $s$  in the collapsed polymer are, for a certain choice of units, expressible as
 \cite{inny,strungout,emerge}
$
\rho\;=\;\frac{1}{2} \frac{1}{g^2 r^2}
\label{eos1}
$ and
$
s\;=\;\frac{1}{g^2 r}
\label{eos2}
$,
with neither of the relations
depending on $D$. Note that, here, $r$ is the radial coordinate for  a fiducial
coordinate system.

Writing  the entropy density  as a function of the energy density,
$
\label{spoly}
g^2 s\; =\; \sqrt{2 g^2 \rho}
$,
$
\frac{1}{T}\;=\;\frac{\partial s}{\partial \rho}= \frac{s}{2 \rho}
$
follows,
which leads, as previously discussed, to
$
sT\;=\;\;2\rho
$.
The pressure can now be directly evaluated,
$
p_r\; =\; sT-\rho\; =\; +\rho\;.
$

And so  $\;p_r=+\rho\;$, as could have been  anticipated \cite{AW}. This ``internal''  equation of state reveals that signals propagate at the speed of light along a closed string. It is not, however, a statement about the interior geometry, as a semiclassical description of the metric is invalidated by strong quantum fluctuations \cite{inny,noclass}.

It should be reemphasized that our negative-pressure solution and the positive-pressure polymer model are meant to be different perspectives of the very same system.
The source of this dichotomy is simple: To avoid the fate of gravitational collapse, one must reject either   a classical geometry or a large  entropy
for the BH interior.
On the other hand, the negative-pressure solution is {\em not} meant to be a
low-entropy or a low-temperature description of a string state. It is rather the picture of the interior according to an external  observer who insists on a non-collapsing, regular geometry under the rules of strictly classical GR. This picture is necessarily
different from that of another external observer who attributes the regularity and
stability against collapse to exotic, maximally entropic matter.  The fact that
two external observers can diagree in this way can be traced to
the inaccessibility of the interior region and can be  viewed as 
a consequence of Hawking's   so-called principle of ignorance \cite{info}.
This disagreement might  also be viewed as a novel form
of observer complementarity \cite{BHfollies}.

\section{Conclusion}

The moral of this paper is as follows: If one considers classical GR, large negative pressure
is, in all likelihood,  a necessary condition for avoiding gravitational collapse. A similar conclusion was reached (and then abandoned) long ago by Bondi \cite{bondi} and more recently revisited in \cite{MM} and many additional works reviewed in \cite{splucc,frolovrev}. We illustrated the idea with our model for the BH as a bound state of closed strings in the Hagedorn phase.

We have seen, however, that if the entropy of the stringy bound state is taken into account, the pressure is positive. Given this discrepancy, how does one know that both descriptions apply to
the same matter system? The answer is simple: These are both describing a compact object whose
every spherically concentric layer behaves just like a BH horizon. This follows from the GR perspective from the perceived form of the inside metric, $\;g_{tt}=g^{rr}=0\;$, and from the stringy perspective
by virtue of the area law for entropy being saturated throughout, $\;s(r)r^d\sim r^{d-1}\;$, and  radial propagation at the speed of light.

The observation that negative pressure need not be associated with a tensile material could have far-reaching implications, as the meaning of  $\;p<0\;$ in a solution to  Einstein's equations is no longer certain. For instance, the accelerated expansion of the Universe may not be due to a  negative-pressure fluid --- the so-called dark energy ---  after all.  The identity $\;p=-\rho\;$ may instead  be a relation  that is used if only classical GR is allowed to explain the accelerated expansion. On the other hand, if some contribution to the entropy is ignored,  there might be others who attribute the expansion of spacetime to matter with a positive pressure.

\begin{acknowledgments}

We would like to thank Emil Mottola for explaining his work to us and critically
reading ours. We also thank Raul Carballo-Rubio for some very useful comments and the numerous colleagues who encouraged us to pursue and answer the question in the title.
The research of AJMM received support from an NRF Incentive Funding Grant 85353, an NRF Competitive Programme Grant 93595 and Rhodes Research Discretionary Grants. The research of RB was supported by the Israel Science Foundation grant no. 1294/16. AJMM thanks Ben Gurion University for their  hospitality during his visit.

\end{acknowledgments}

\end{document}